\documentclass{elsart}

 \usepackage{epsfig}

\usepackage{amssymb}

\begin{document}

\begin{frontmatter}

 \title{Search for possible neutrino radiative decays during the 2001
 total solar eclipse}

 \author[bof,tesre]{S. Cecchini}
 \author[bof]{D. Centomo}
 \author[bof]{G. Giacomelli}
 \author[bof]{R. Giacomelli}
 \author[bof,iss]{V. Popa}
 \author[bof,iss]{C.G. \c{S}erb\u{a}nu\c{t}}
 \author[bof,per]{R. Serra}
\address[bof]{Dipartimento di Fisica dell'Universit\`{a} and INFN Sezione di Bologna,
 I-40127 Bologna, Italy}
\address[tesre]{IASF/CNR, I-40129 Bologna, Italy}
\address[iss]{Institute for Space Sciences, R-77125 Bucharest M\u{a}gurele, Romania}
\address[per]{Osservatorio Astronomico, I-40017 San Giovanni in Persiceto, Italy}

\begin{abstract}
We present the results of the observations performed in the occasion of the
June 21$^{st}$, 2001
total solar eclipse, looking for visible photons emitted through a possible
radiative decay of solar neutrinos.  We establish lower limits for
the $\nu_2$ or $\nu_3$
 proper lifetimes
$\tau_0/m \geq 10^3$ s/eV, for neutrino masses larger than $10^{-2}$ eV.

\end{abstract}

\end{frontmatter}

\section{Introduction}
\label{intro}

In the last few years it has become  clear that neutrinos have non-vanishing
masses, and that the neutrino flavor eigenstates ($\nu_e$, $\nu_\mu$ and $\nu_\tau$)
are superpositions of mass eigenstates ($\nu_1$, $\nu_2$ and $\nu_3$),
giving rise to neutrino oscillations. For  recent
reviews, see \cite{gg,pak}. In this context, neutrinos can
undergo radiative decays, e.g. $\nu_2 \rightarrow
\nu_1 + \gamma$, as initially suggested in \cite{sciama}; the first searches
for such decays were based on astrophysical considerations (see eg. \cite{cow}). The
status of the decaying theory and phenomenology was summarized in \cite{sciama2}.

Radiative decays are allowed if
the involved neutrinos have a non-vanishing  magnetic dipole moment; the
very stringent existing experimental limits ($\mu_\nu
< 1.3 \times 10^{-10} \mu_B$
\cite{pdg02}) refer to the neutrino flavor eigenstates
and  are not directly applicable to possible dipole magnetic moments of neutrino mass
eigenstates.

Neutrino decays (radiative or not) have been searched for  indirectly (from
astrophysical arguments such as Supernova physics or the absence of $\gamma$ rays
in the Sun radiation, or from IR background measurements), by re-interpreting solar
and atmospheric neutrino data from large experiments,
as well as directly (experiments near nuclear reactors,
analysis with cosmic ray detectors or during total solar eclipses). All published
searches yield lower limits for the neutrino lifetimes, which are strongly
sensitive to the assumed neutrino mass and mass hierarchy scenario.

The astrophysical neutrino lifetime lower limits are usually large (e.g. $\tau_0/m
> 2.8 \times 10^{15}$ s eV$^{-1}$ where $\tau_0$ is the lower proper lifetime
limit for a neutrino of mass $m$, \cite{blud}), but they are indirect and
rather speculative limits.

Much lower ``semi-indirect" limits were deduced from the re-interpretation of solar and
atmospheric neutrino data. Earlier attempts to explain the solar neutrino or
atmospheric neutrino anomalies only in terms of neutrino decay have been ruled out
by the existing evidence \cite{out}; the present accepted
explanations are based on neutrino oscillations, but
do not exclude the hypothesis of neutrino decays.
As an example,  from the SNO data \cite{sno1,sno2} a proper lower limit of
$\tau_0/m > 8.7 \times 10^{-5}$ s eV$^{-1}$ was deduced \cite{ab}. By analyzing
 all available solar neutrino data, other limits were obtained:
$\tau_0/m > 2.27 \times 10^{-5}$ s eV$^{-1}$ for the MSW solution,
and $\tau_0/m > 2.78 \times 10^{-5}$ s eV$^{-1}$ for the vacuum oscillation solution of
the
solar neutrino problem (SNP) \cite{aj}, or, following a different approach, $\tau_0/m
> 10^{-4}$ s eV$^{-1}$ \cite{jb}.

Direct searches for radiative neutrino decays have been also performed. As an example
we quote here the search for decay photons in the visible spectrum performed in
the vicinity of a nuclear reactor \cite{reactor}, yielding $\tau_0/m$ lower
limits in the range  $10^{-8}$ to nearly 0.1 s eV$^{-1}$, assuming neutrino
relative mass differences $\Delta m / m$ between 10$^{-7}$ and 0.1.
 Recently,
a search for $\gamma$ photons, using the Prototype
Borexino Detector at Gran Sasso \cite{borex} reported $\tau_0/m$ lower limits
of $1.5 \times 10^3$ s eV$^{-1}$ (assuming a polarization parameter $\alpha =
-1$ for the parent neutrino), $ 4.4 \times 10^3$ s eV$^{-1}$ (for $\alpha =0$) and
$9.7 \times 10^3$ s eV$^{-1}$ (for $\alpha =+1$).

In a pioneering experiment  performed during the
Total Solar Eclipse (TSE) of October 24, 1995
a search was made for visible photons emitted through  radiative decays
of solar neutrinos during their flight between the Moon and the Earth \cite{vanucci}.
 In their analysis,
  the authors assumed
  that all neutrinos  have masses of the order of
few eV,   $\Delta m_{12}^2 = m_2^2 - m_1^2
 \simeq 10^{-5}$ eV$^2$, and an average  energy of
860 keV. Furthermore they assumed that all decays would lead to visible photons,
which would travel nearly in the same direction as the parent neutrinos, thus
leading to a narrow spot of  light
 coming from the direction of
 the center of the dark disk of the
Moon. In the absence of a positive signal, this search yielded a lower proper lifetime
 of 97 s, for neutrinos with $m_\nu$ ~ few eV \cite{vanucci}.

Some of the authors of this paper
 were involved in two similar experiments, during
the total solar eclipses of August 11, 1999 (in Romania)
\cite{n1,n2}, and of June 21, 2001 (in Zambia) \cite{n3,n4}.

In 1999 the bad weather conditions did not allow the planned observations with the properly
designed experimental apparata; but we could analyse  the images of a videotape recorded
by a local Romanian television (R\^{a}mnicu V\^{a}lcea).
The study of
the data was performed
in the hypothesis of a possible decay
\begin{equation}
\label{deca}
\nu_2 \rightarrow \nu_1 + \gamma,
\end{equation}
with $m_2 > m_1$, and for
two values of $\Delta m^2$  suggested by the
MSW SMA (Small Mixing Angle) and LMA (Large Mixing Angle) solutions of the
 SNP, allowed by the then available  data from solar
neutrino experiments.

We developed a Monte Carlo (MC) simulation of radiative
solar neutrino decays, considering the neutrino energy spectrum predicted
by the Standard Solar Model (SSM) \cite{bahcall1} and the mass of the $\nu_1$
in the range of 1 - 10 eV, as it was expected at that time.
 Since the angular resolution of the data was not very
good,
we considered the Sun as a pointlike  source. The simulation has shown that
the expected signal would be a narrow spot of light in the direction of the center
of the Sun, and allowed an evaluation of the fraction of decays yielding visible
photons as function of the chosen neutrino $\nu_1$ mass, $m_1$,
and $\Delta m^2$ values. The lower limits obtained from the 1999 TSE
range between $1.8 \times 10^{-2}$ s eV$^{-1} < \tau_0/m <  14.5$ s eV$^{-1}$.

It may be worth to mention that the Particle Data Group best fit lower limit for the
neutrino lifetime, in the printed version of 2002\, was $7 \times
10^{9}$ s eV$^{-1}$, while in the 2003 partial upgrade the best fit
is only 300 s eV$^{-1}$ \cite{pdg02}.

In this paper we present the results obtained from the analysis of our 2001 TSE
observations.
This experiment
allowed the collection of better resolution data, so the real
spatial distribution of
the solar neutrino yield had to be considered. The recent SNO
results \cite{sno1,sno2}  favor  the LMA solution  and
 indicate also the presence of a $\nu_\tau$ component in the solar neutrino
flux at the Earth level. Furthermore, they indicate a possible lower limit for the
sum of the neutrino  masses of about $4.8 \times 10^{-2}$ eV.
The WMAP (Wilkinson Microwave Anisotropy Probe)
results after the first year flight, assuming three degenerate neutrino species
\cite{map} limit the mass of the neutrino to $\leq  0.23$ eV ( 95\%
CL).

The new Monte Carlo simulation code
developed for the analysis of our 2001 data was described elsewhere
\cite{nou}. It is based on the latest updates of the Standard Solar Model
\cite{bahcall2} and on a full 3-D geometry of the solar neutrino production, decay and
photon detection, as initially proposed in \cite{frere}.

\section{Experimental data}

The experimental data used for this search consist of two sets of digital images,
obtained with two different instruments and
 different exposures, magnifications and resolutions, during the
 TSE of June 21$^{st}$ 2001. For the observation we have chosen a
location 50 km North of Lusaka, Zambia, in the vicinity of a farm that could provide
 electrical power. The site was located at 14$^o$56' lat. S, 28$^o$14'
long. E, and at an altitude of about 1200 m a.s.l. The distance to the centrality
line was approximately 8 km.

One data set (referred to as ``set A") consists in 4149
frames obtained from a digital film made with a
video-camera equipped with a $2 \times$ telelens and a $10 \times$ optical zoom. The
recording rate was 25 frames per second. The field of view of each pixel
of the CCD camera was 9.97" $\times$ 10.74" (arcsec).

The second data set (that will be referred to as ``set B") consists in 10 digital
photographs made with a Matsukov - Cassegrain telescope (90 mm $\phi$ and f = 1250 mm),
coupled to a digital camera.
The camera was set to emulate a photographic film of 400 ASA sensitivity.
The exposure time for
each image was 2 s, and the digital camera automatically adjusted the
aperture as function of the luminosity of the solar corona. We took into consideration
this effect in our analysis. In order to avoid systematic effects due to the optical
system
or to the CCD, we slightly changed the orientation of the telescope from one
frame to another.
The field of view of each pixel of the CCD
was 1.14" $\times$ 1.14".

For both data sets and for each image,
we determined the center of the dark disk of the Moon.
The signal of a possible radiative decay of solar neutrinos during
their flight from the Moon to the Earth should be correlated to the position of
the center of the Sun.
The totality phase of the TSE at the observation point  lasted about
$t_{tot.} = 3.5$ minutes,
so the center of the Sun continuously changed its position with respect to the center
of the dark disk of the Moon. In equatorial coordinates, the
velocity components (declination and ascension) of the relative movement of the
Moon center with respect to the Sun center were: $v_\delta \simeq 0.055$"/s and
$v_\alpha \simeq 0.758$"/s. As the angle between the ``horizontal"  x-axis of our
images and the ecliptic was about 53$^0$, we computed the ``x" and ``y" components
of the relative velocity of the
center of the Moon with respect to the Sun center: $v_x \simeq 0.455$"/s and
$v_y \simeq 0.033$"/s. The distance (in arcsec) between the two centers
was computed for any time $t$ from the start of the totality,
 knowing that at $t = t_{tot.}/2$ the Moon and the Sun  centers were
coincident. This allowed us to know, for each image, the shift (in pixels) between
the Moon center and the center of the Sun, and thus the displacement
to apply to the images before adding one to another.

The procedure to search for possible signals in each data set is based on the
wavelet decomposition of each of the two total
images, and we retained as useful data only
the larger dyadic squares around the center of the Sun. In the case of  set
A, the dimension of this square is $64 \times 64$ pixels$^2$, while for set B
is $512 \times 512$ pixels$^2$.
Fig. 1 illustrates this procedure for a typical image of data
set B. The border of the dark Moon disk is obtained
by requiring a variation of 100 Aquisition Digital Units
(ADU) from a pixel to the next one. The resulting
shape is then fit to a circle.  The border pixels and the fit circle are
shown in Fig. 1. The position of the center of the Sun with respect to the
center of the Moon is calculated as a function of the time at which the photo
was made; it is represented in Fig. 1 by the small cross in the middle of
the image. The $512 \times 512$ pixels$^2$ area, centered on the center of the Sun,
is then obtained for each picture (the solid rectangle in Fig. 1);  all the
similar zones of the 10 pictures are summed  toghether for further processing.
The relative displacement of the Sun center from the beginning to the end of
data taking is indicated in Fig. 1
by the  small points at the center of the field;
 a $512 \times 512$ pixels$^2$ rectangle centered on the center
of the Moon is also drawn (the dotted square).
\begin{figure}
\vspace{-10mm}
\begin{center}
       \mbox{  \epsfysize=12cm
               \epsffile{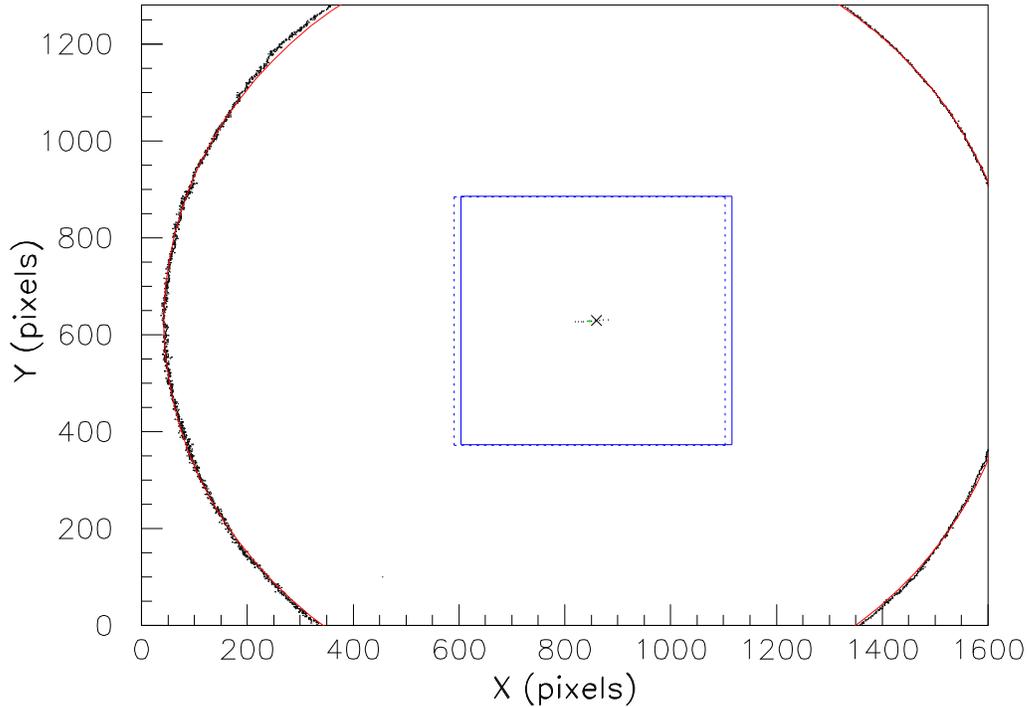}}
\caption{Illustration of the procedure used to select the largest dyadic square
centered on the center of the Sun from the pictures of set B
obtained during the TSE. See
text for details.}
\end{center}
\end{figure}
This procedure  was applied to all digital video frames of data set A
and to all digital photographs of data set B; the sums were then computed for each
color channel (red, green and blue) and for their sum (the ``white" channel).

\section{Data analysis}

The MC simulations of the solar neutrino radiative decays \cite{nou},
$\nu_2 \rightarrow \nu_1 + \gamma$, assuming $\Delta m^2_{21} = 6 \times 10^{-5}$
eV$^2$, $10^{-3} \leq m_1 \leq 3 \times 10^{-1}$ eV,
indicate that the expected signal in the optical band should consist in a narrow
luminosity peak (about 50" wide) in the direction of the center of the Sun. If the
$\nu_3$ mass eigenstate is also present in the solar neutrino flux at the Earth, the
$\nu_3 \rightarrow \nu_{2,1} + \gamma$ decays,
with $\Delta m^2_{32(1)} = 2.5 \times 10^{-3}$
eV$^2$, would yield a broader signal (about 250"), possibly showing
some ``rings" at
 200 - 300 arcseconds, concentric to the position of the center of the Sun.
Most of the background light (diffraction of the coronal light on the borders of the
Moon, diffuse sky light, ashen light (light reflected by the Earth on the surface of the
Moon), etc.), are characterized by larger angular scales. Thus the wavelet decomposition
of the compound images  is a
proper tool to search for a possible decay signal.

 We used the
simple Haar wavelet basis \cite{wave}.
The $n$-order term of the decomposition is obtained by dividing the $N \times N$
pixels$^2$ image in square fields of $N/2^n \times N/2^n$ pixels$^2$ and averaging the
luminosity in each field; the averages are then removed and the resulting image,
the $n$-order residual, can be used to obtain the $(n+1)$-order term. Thus, each
decomposition term results in an image in which objects of the corresponding
scale are dominant, while the residuals contain  information for
smaller dimension scales.

The decay signal is searched for by averaging the luminosity of the images over
``rings" centered on the position of the center of the Sun. As the wavelet analysis
requires a dyadic dimension of the field (the number of pixels on each border
of the image is a power of 2), there is no  ``central pixel"; so we have
considered each of the four pixels adjacent to the image center as ``central" and
then averaged the obtained luminosity profiles.

Fig. 2 shows the luminosity profiles obtained
from the total images A (Fig. 2a) and B (Fig. 2b).
Both graphs refer to the ``white" channels, that is the sum over the three color
channels. The vertical scales are expressed in ADU;
the meaning of 1 ADU (different for each instrument)
 is discussed in Section 4.
\begin{figure}
\vspace{-10mm}
\begin{center}
       \mbox{  \epsfysize=8.3cm
               \epsffile{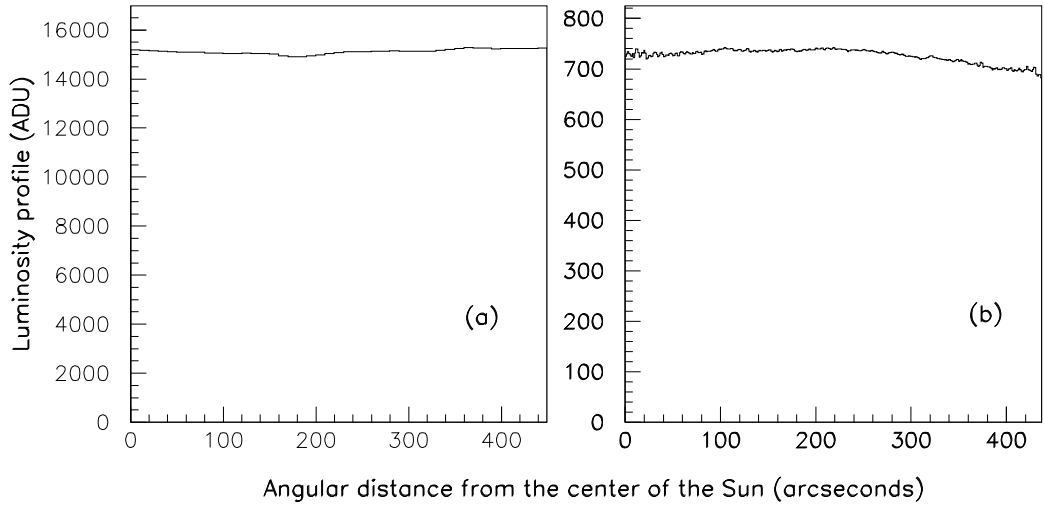}}
\caption{Luminosity profiles (in white light) for (a) the compound image A
and (b) B (raw data).}
\end{center}
\end{figure}
The  difference between the two total images is due to the different CCD
sensitivities, spatial resolution and optical feathures of the instruments.
Data set A presents
no clear structure, data B might contain some at relatively large $\theta$.

Most of the background is removed by
the wavelet decomposition. Fig. 3 shows the  residuals
of order 5 for the
data A, and of order 8 for the B data,
in the
``white" channel.  In these residuals (the highest possible order for each data sets),
 the contributions
of  structures larger than $2 \times 2$ pixels$^2$ are removed.
\begin{figure}
\vspace{-10mm}
\begin{center}
       \mbox{  \epsfysize=8.5cm
               \epsffile{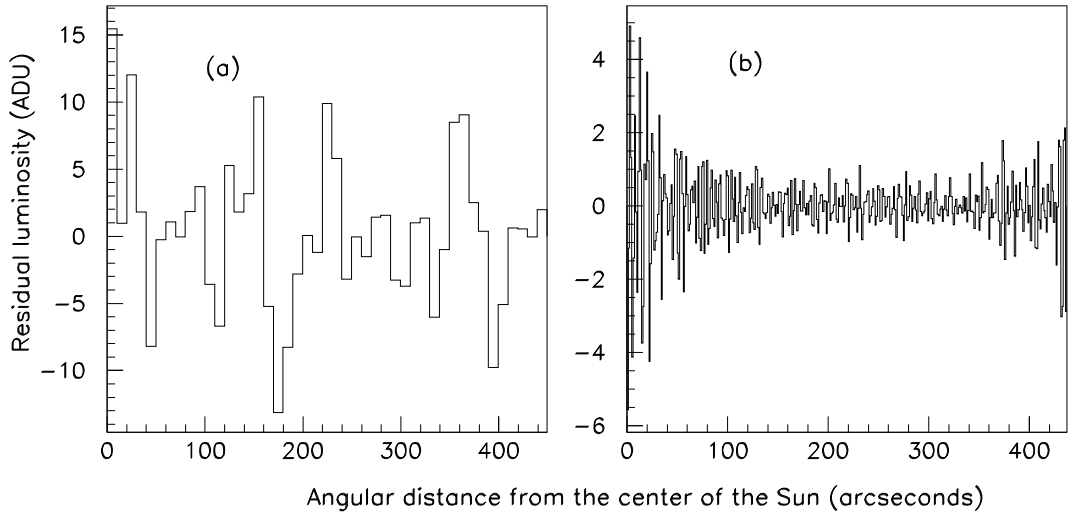}}
\caption{Luminosity profiles of the last residuals from the wavelet
decomposition of (a) the compound images A -fifth order-, and (b) B
-eight order-. Both profiles are obtained in the ``white" channel.}
\end{center}
\end{figure}
Both Figs. 3a and 3b are dominated by fluctuations, so we may conclude that at our
experimental sensitivities, data sets A and B do not contain indication
for a narrow
light excess from the direction of the center of the Sun. The increase of the
fluctuation amplitudes near the limits of the angular range, clearly
noticeable in Fig. 3b, is natural, as the number
of pixels that contribute to the average luminosity in that regions is smaller than
in the rest of the image.

We recall that maxima confined to the first bin and present in all color channels were
observed in the TSE 1999 data \cite{n2,n3,n4}. Probably we
observed  the so-called ``Poisson spot" consisting in a
peculiar diffraction phenomenon that is produced when a circular opaque object,
with size much smaller than the distance to a screen, but much larger than the
light wavelength, is illuminated by a parallel beam \cite{pois}.
 In such  conditions, a point-like
spot of light could be observed on the screen, in the middle of the shadow of the object.
In 1999 we  analyzed  a video  filmed
 close  to the point of maximum eclipse,  when the center of the Sun,
the center of the Moon and the observer were in very good alignment.

Since in 2001 we were in a location
8 km away from the line of centrality of the TSE,
 one dos not expect that the same effect
is seen.
Such signal could be enhanced by aligning the 10 pictures with respect to
the center of the Moon. Fig. 4 shows the results of such an analysis.
\begin{figure}
\vspace{-10mm}
\begin{center}
       \mbox{  \epsfysize=8.5cm
               \epsffile{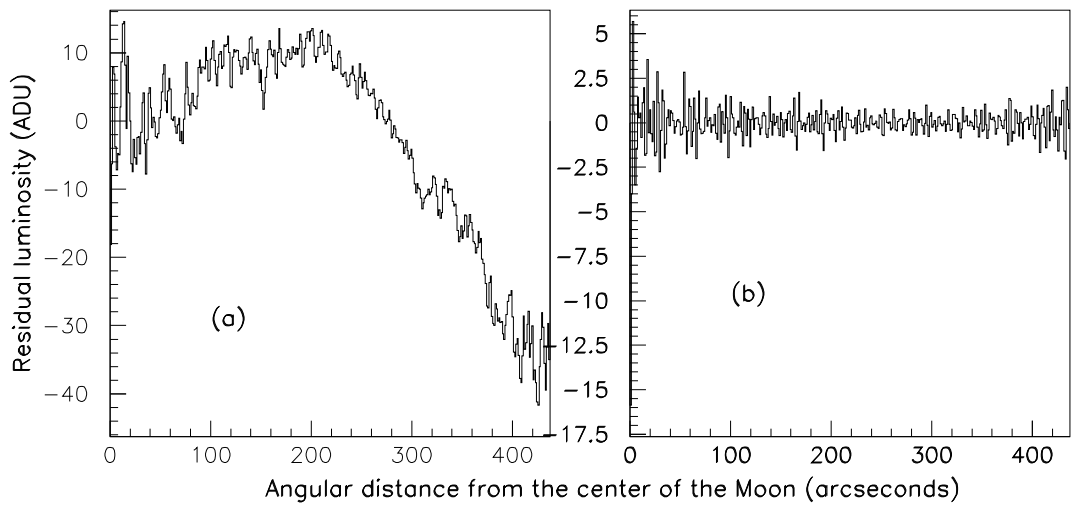}}
\caption{``White" light
luminosity profiles obtained by aligning the images of data set B with
respect to the center of the Moon, from (a) the 0 order (thus after removing
the average luminosity) and (b) the
$8^{th}$ order residuals}
\end{center}
\end{figure}
In Figs. 4a and 4b there is no peak associated with the center of the Moon disk. A
darker spot is instead seen in the center of the Moon; it
should correspond to the dark Schr\"{o}ter
area.
The  structures in Fig 4a may be produced by the image of the Moon surface in the light
of the Sun reflected by the Earth.  Their
absence in Fig. 4b is due to the rejection of large structures by the
wavelet decomposition.

The expected signal from a decay (1), considering
its MC estimated width \cite{nou},
should be better seen in the fourth order wavelet term
of data sets A and B, as it corresponds to structures with about 40" - 60" width.
Fig. 5 shows the luminosity distributions for this term. Note that
each bin is an average over $4 \times 4$ pixels in the case of data set A (Fig. 5a), and
over $32 \times 32$ pixels for set B (Fig. 5b).
\begin{figure}
\vspace{-6cm}
\begin{center}
       \mbox{  \epsfysize=15cm
               \epsffile{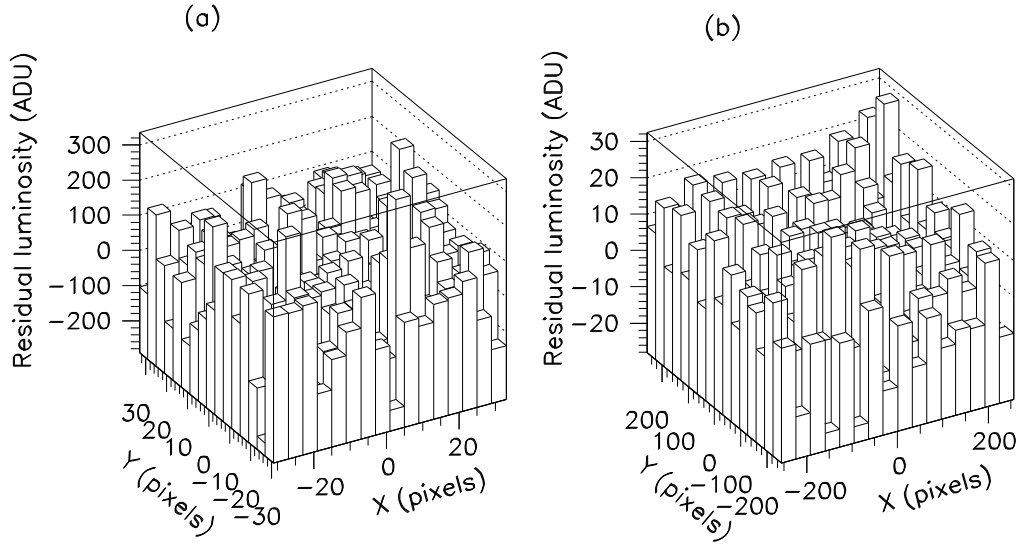}}
\caption{``White" light luminosity distributions of the fourth order wavelet term of (a)
the sumed images
A and (b) B  (centered on the Sun).}
\end{center}
\end{figure}
No central maximum is present in both data sets, and we can use these wavelet terms
to determine lower lifetime limits for the investigated decay.

The search for the $\nu_3 \rightarrow \nu_{2,1} + \gamma$ decay is more difficult, as
the expected signal is broader (about 200" - 300") \cite{nou}.
Since the wavelet decomposition could remove
broad signals, we searched for the corresponding signal in the raw data. Data set A  does
not exhibit such a signal (Fig. 2a), but the luminosity curve obtained from set B (Fig. 2b)
could suggest the presence of the expected signal. Considering the good resolution and
sensitivity of instrument B, we assumed that the ashen light (the luminosity of the
Moon surface due to the light of the Sun reflected by the Earth) could be responsible
for most of the signal seen in B.
In order to verify that, we performed the same analysis on an image of the full Moon, obtained
with the same telescope and camera. The results of this test are shown in Figs. 6a and
6b.
\begin{figure}
\vspace{-10mm}
\begin{center}
       \mbox{  \epsfysize=8.5cm
               \epsffile{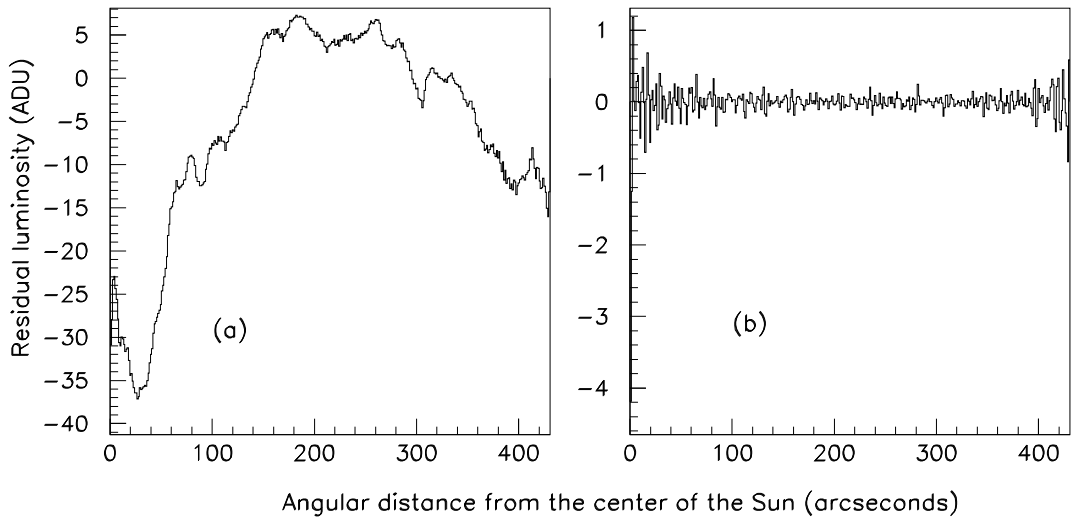}}
\caption{``White" light luminosity profiles obtained from an image of the full Moon
 from (a) the 0 order and (b) the $8^{th}$ order residuals.
  The Moon image was obtained with the same equipment used for taking the set B data.}
\end{center}
\end{figure}
The structures in Fig. 6a are  similar to those in Figs. 2b and 4a;  we should take into
consideration that the
Earth reflects the light of the Sun as a convex mirror, thus the central part of the Moon
receives more light from the Earth than the rest of it. Instead the Sun iluminates the
Moon uniformely. This observation suggests that we cannot simply remove the image of the
full Moon from the data, as it would create a fake signal in the central part of the
resulting image. Thus, one should develop a reliable model of the ashen
light; alternatively one should use the wavelet decomposition.
 Note  that the
exposure conditions for the image of the full Moon were different than those during
the eclipse, so the ADU values are not directly comparable.
As we cannot determine which is the real contribution of the ashen light in set B, we
can use the results only to determine a lower limit for the $\nu_3$ lifetime.

\section{Calibrations}

The luminosity values given in the previous Sections in ADU
 have to be converted   in numbers of
photons hitting each pixel of the CCD's of our detectors. For this purpose
two sets of calibrations
were made, using the monochromators of the Catania and Bologna Astronomical Observatories.
The obtained calibration
curves are consistent one with the other. We took care to calibrate the instruments in
conditions as close as possible to the conditions in which the TSE data were recorded.
The digital videocamera was operated in the same conditions as during the eclipse;
the pictures  with the
digital camera  were taken with exposures  of
2 s, and a fixed diaphragm aperture (f/2.5). As mentioned in Section 2, the
10 digital pictures in data set B were obtained with different apertures; the total
20 s of exposure in the eclipse conditions is equivalent to about 4.7 s
 in the calibration conditions.

Fig. 7 shows the
calibration curves obtained.
In the vertical scales are indicated the number of incident photons that
produce 1 ADU in one pixel of the CCD's. Each curve corresponds to one color channel.
\begin{figure}
\vspace{-6cm}
\begin{center}
       \mbox{  \epsfysize=15cm
               \epsffile{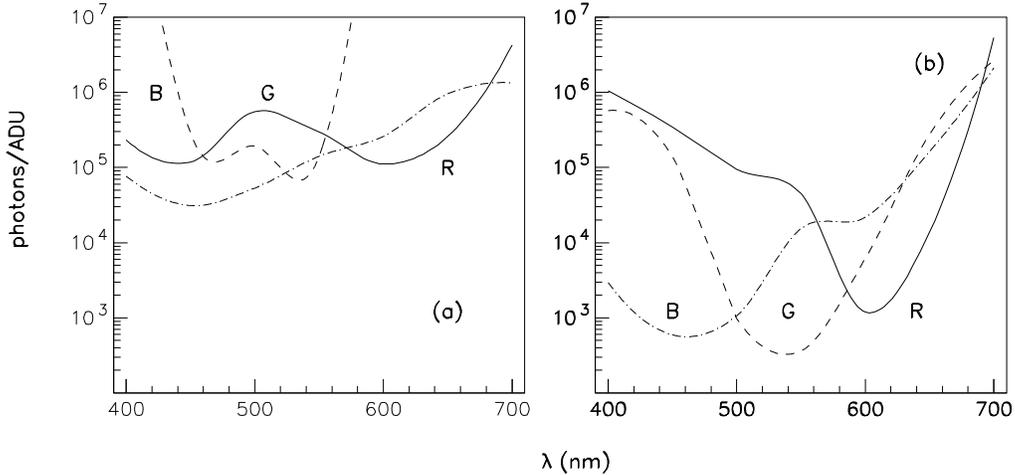}}
\caption{Calibration curves (number of incident photons that produce one ADU
plotted as function
of the wavelength): (a) for the digital videocamera, and (b)
the digital camera with telescope.
 The results concerning the Red, Green and Blue channels are represented as
solid, dashed and dot-dashed curves, respectively.}
\end{center}
\end{figure}
From these calibrations it follows that the wavelength averaged
number of visible photons that
produces one ADU in the ``white" channel of  data set A is about $7.3 \times 10^4$,
while for  data set B is two orders of magnitude smaller: $8.9 \times 10^2$. This
could explain why we may see the ashen light only in data B.

\section{Results and discussions}

 Both data  A and B were processed in order to search for signals produced by a
possible radiative decay of the heavier component of solar neutrinos. In the absence of
a valid signal, we used the two sets to obtain lower limits for the neutrino
radiative decay lifetime.

Let us assume the
simplest scenario
in which the Sun electron neutrinos are a superposition of only two mass eigenstates
$\nu_1$, $\nu_2$, with
a mixing angle $\theta_{12}$
\begin{equation}
| \nu_e \rangle = |\nu_1 \rangle \cos \theta_{12} + |\nu_2 \rangle \sin \theta_{12},
\end{equation}
Using the calibrations of our instruments, we  transform the ADU
 into numbers of
photons $N_\gamma$, that produced the analyzed images.
The lifetimes of the $\nu_2 \rightarrow \nu_1 + \gamma$ decay
 in the laboratory
frame of reference, can be calculated from the standard deviations of the fourth order
terms of the wavelet decomposition of our data
\begin{equation}
N_\gamma = P \Phi_2 S_M t_{obs} \left(1-e^{-\frac{\langle t_{ME} \rangle}{\tau}}
 \right)
e^{-\frac{t_{SM}}{\tau}}~,
\end{equation}
where
 $P$ are the mass - dependent probabilities estimated by the MC simulation
 \cite{nou},
  $\Phi_2 = \Phi_\nu \sin^2 \theta_{12}$, ($\Phi_\nu$ is the
flux of solar neutrinos at the Earth (or Moon) and $\theta_{12}$ the mixing angle)
is the local flux of solar $\nu_2$ mass eigenstate neutrinos,
$S_M$ is the area of the
Moon surface covered by the analysis (the area of the square on the Moon that
is visualized inside the dyadic square) and $t_{obs}$ is the time of observation.
$\langle t_{ME} \rangle$ is the average
time spent by solar neutrinos inside the observation cone
(about one third of the flight
time from the Moon to the Earth), and $t_{SM}$ is the time of flight of the neutrinos
from the Sun to the Moon \cite{nou}.

The MC probabilities $P$ are shown in Figs. 8a,b, for
data sets A and  B, respectively. Assuming that
a neutrino decay occurs during the flight of solar neutrinos between the Moon
and the Earth, inside the region of space ``visible" by the experiment,
and that the emitted photon reaches the detector,
the probability $P$ includes
 kinematic effects (dominated by the assumed polarization $\alpha$ of the $\nu_2$
beam), the request that the emitted photon should be in the visible energy range and
the ``imprints" of the initial neutrino characteristics (energy and ``birth" position),
 according to the
SSM \cite{bahcall2}.
\begin{figure}
\vspace{-6cm}
\begin{center}
       \mbox{  \epsfysize=15cm
               \epsffile{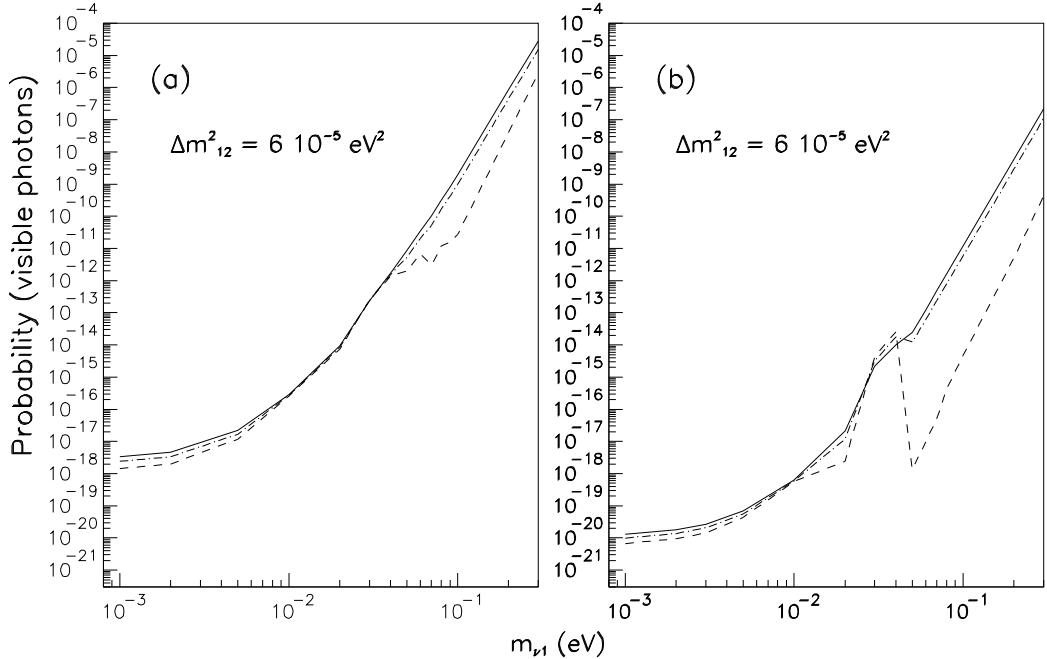}}
\caption{Monte Carlo Probabilities for the production and arrival to the detector of the decay
photons resulting from $\nu_2 \rightarrow \nu_1 + \gamma$ transitions during the
neutrino flight from the Moon to the Earth. The simulations were made for
(a) data set A and (b) set B. The solid, dot-dashed and dashed lines
correspond to three different neutrino polarizations, $\alpha = -1$, 0 and +1,
respectively.}
\end{center}
\end{figure}
The simulations were made for three possible values of the polarization of
the decaying neutrinos: $\alpha = -1$, which corresponds to Dirac left-handed neutrinos
(like for the massless neutrinos in the Standard Model),
$\alpha = 0$ (Majorana neutrinos),  and for right handed Dirac
neutrinos ($\alpha = +1$).

Similar simulations where made assuming $\nu_3 \rightarrow \nu_{(2,1)} + \gamma$,
with $\Delta m^2_{3,(2,1)} = 2.5  \times 10^{-3}$
eV$^2$. The computed probabilities are presented in Fig. 9.

\begin{figure}
\vspace{-5cm}
\begin{center}
       \mbox{  \epsfysize=15cm
               \epsffile{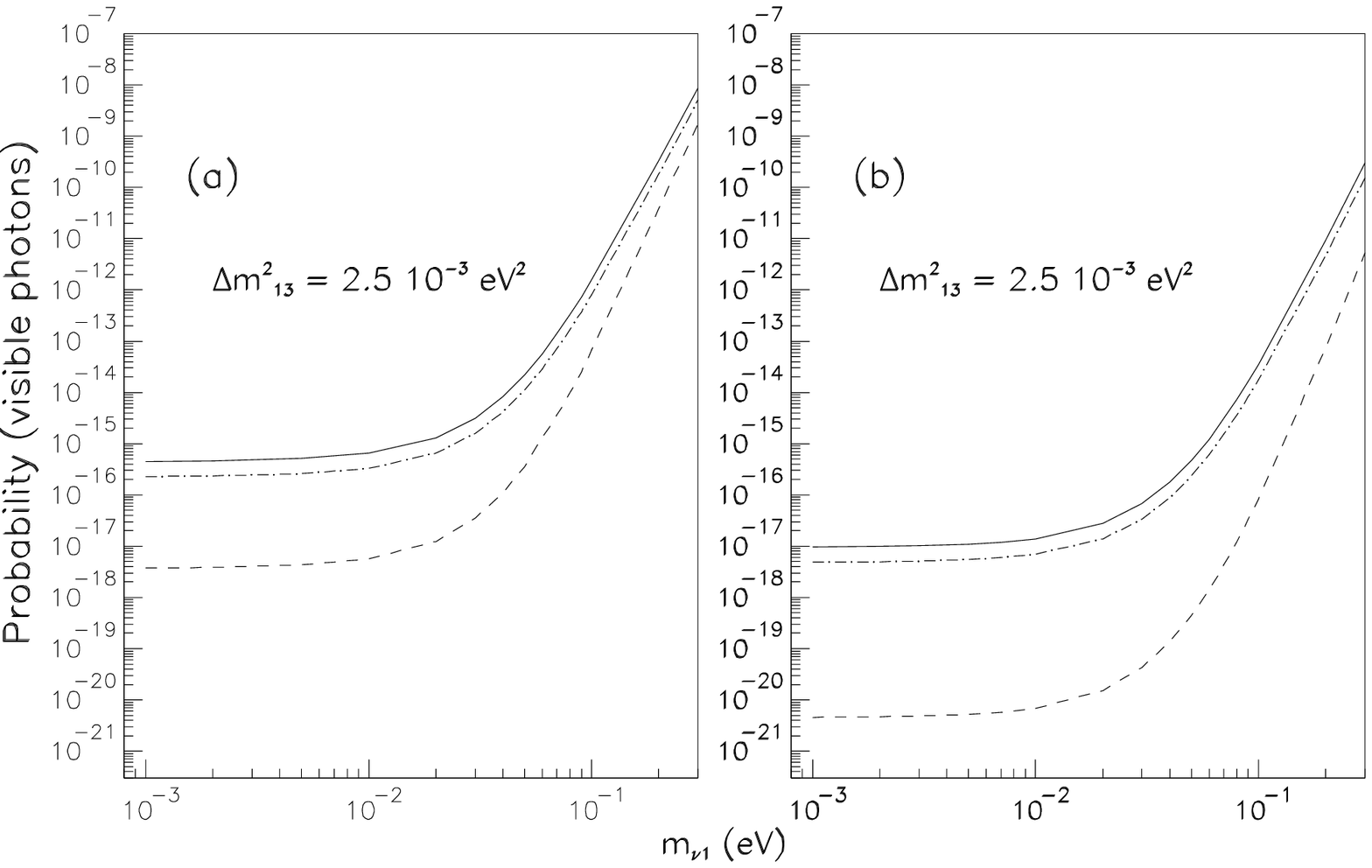}}
\caption{Monte Carlo Probabilities for the production and arrival to the detector of the
decay
photons resulting from $\nu_3 \rightarrow \nu_{(2,1)} + \gamma$ transitions during the
neutrino flight from the Moon to the Earth. The simulations were made in the conditions
of (a) data set A and (b) set B. The solid, dot-dashed and dashed lines
correspond to three different neutrino polarizations, $\alpha = -1$, 0 and +1,
respectively.}
\end{center}
\end{figure}

The 95\% CL lower limits for the $\nu_2$ lifetime, in our case of no signal,
are obtained by the substituting in Eq. 3
$N_\gamma$ with $3\sigma_{N_\gamma}$ of the forth order wavelet terms decomposition
of the data. They are shown with thicker lines
in Figs. 10a (data  A) and 10b (data  B),
assuming that $\nu_2$ is a Dirac (lefthanded or righthanded) or a Majorana neutrino.

The recent limits obtained from the  Borexino
 Counting Test Facility \cite{borex} are also shown, for comparison.
The arrows labelled ``SNO" and ``WMAP" indicate the lower neutrino mass limit
reported by SNO \cite{sno1,sno2}, and the upper mass limit obtained by WMAP \cite{map}.
The limit obtained by the first TSE experiment \cite{vanucci} is indicated by the
horizontal arrow; note that this limit,  obtained using different physical
hypotheses, is valid for neutrino masses of few eV.

\begin{figure}
\vspace{-5cm}
\begin{center}
       \mbox{  \epsfysize=15cm
               \epsffile{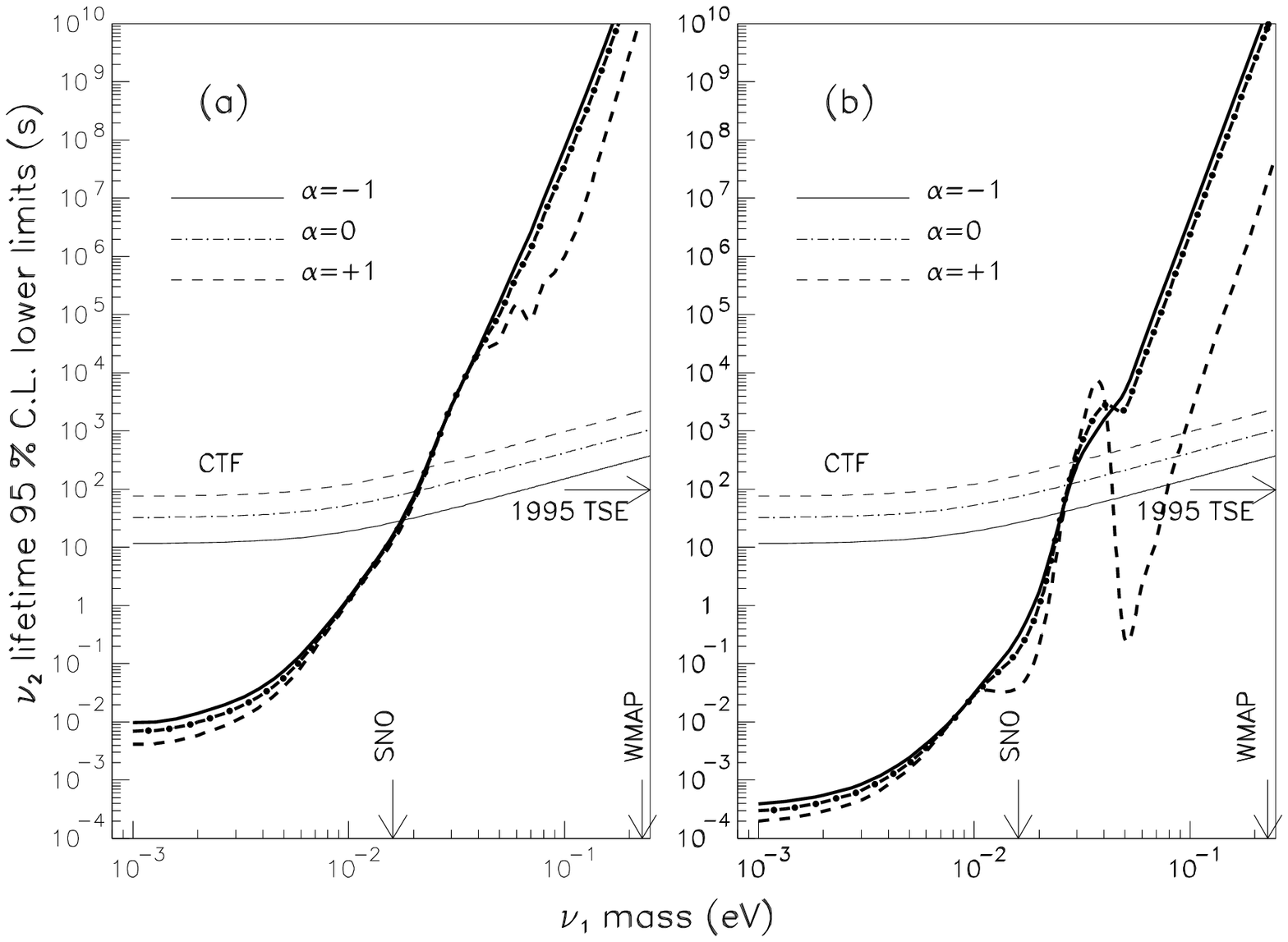}}
\caption{The 95\% CL lower limits for the $\nu_2$ proper lifetime,
as function of the $m_{\nu_1}$, obtained from data sets (a) A and (b) B.
The results are valid in a neutrino mixing scenario with only two
generations, and $\Delta m^2_{2,1} = 6 \times 10^{-5}$ eV$^2$.
The discontinuities in the proper lifetime limits for righthanded neutrinos originate in
the MC probabilities and reflect the changements in the initial neutrino energy imposed by
the condition of obtaining visible decay photons pointing to the Earth.
Other relevant limits are also
indicated (see text).}
\end{center}
\end{figure}

 Neutrino lifetime values larger than our lower limits
 are not in conflict with the oscillation explanation of the
solar neutrino deficit. The neutrino time of flight from the Sun to the Earth is
about 500 s (in the laboratory frame of reference). The Lorentz boost for a solar
neutrino with a mass of about 0.02 eV is $\gamma \simeq 1.5 \times 10^{7}$,
so the fraction of $\nu_2$ that would decay into $\nu_1 + \gamma$, assuming
$\tau_0 \simeq 60$s (in the c.m.) would be only  $\simeq 5 \times 10^{-7}$.

The atmospheric neutrino data indicate  $\nu_\mu \rightarrow \nu_\tau$
oscillations with
 maximal mixing \cite{macro,miri,sk}.  The two
generation mixing of solar neutrinos is an approximation; the ``heavy" component  of
the flux could then be a mixture  of $\nu_2$
and $\nu_3$ mass eigenstates. The  mixing angle $\theta_{13}$ is not yet
measured, but it should be small; we
 assume $\sin^2 \theta_{13} \simeq 0.1$.
Data set A does not contain ``signals"
similar to the MC expectations \cite{nou}, and in data set ``B" the presence of
the ashen light
  creates a luminosity
profile that could mask a possible signal. In both cases, we  obtain the 95\% CL $\nu_3$
lifetime lower limits, from the raw images, shown  in Fig. 11.
\begin{figure}
\vspace{-2 cm}
\begin{center}
       \mbox{  \epsfysize=11cm
               \epsffile{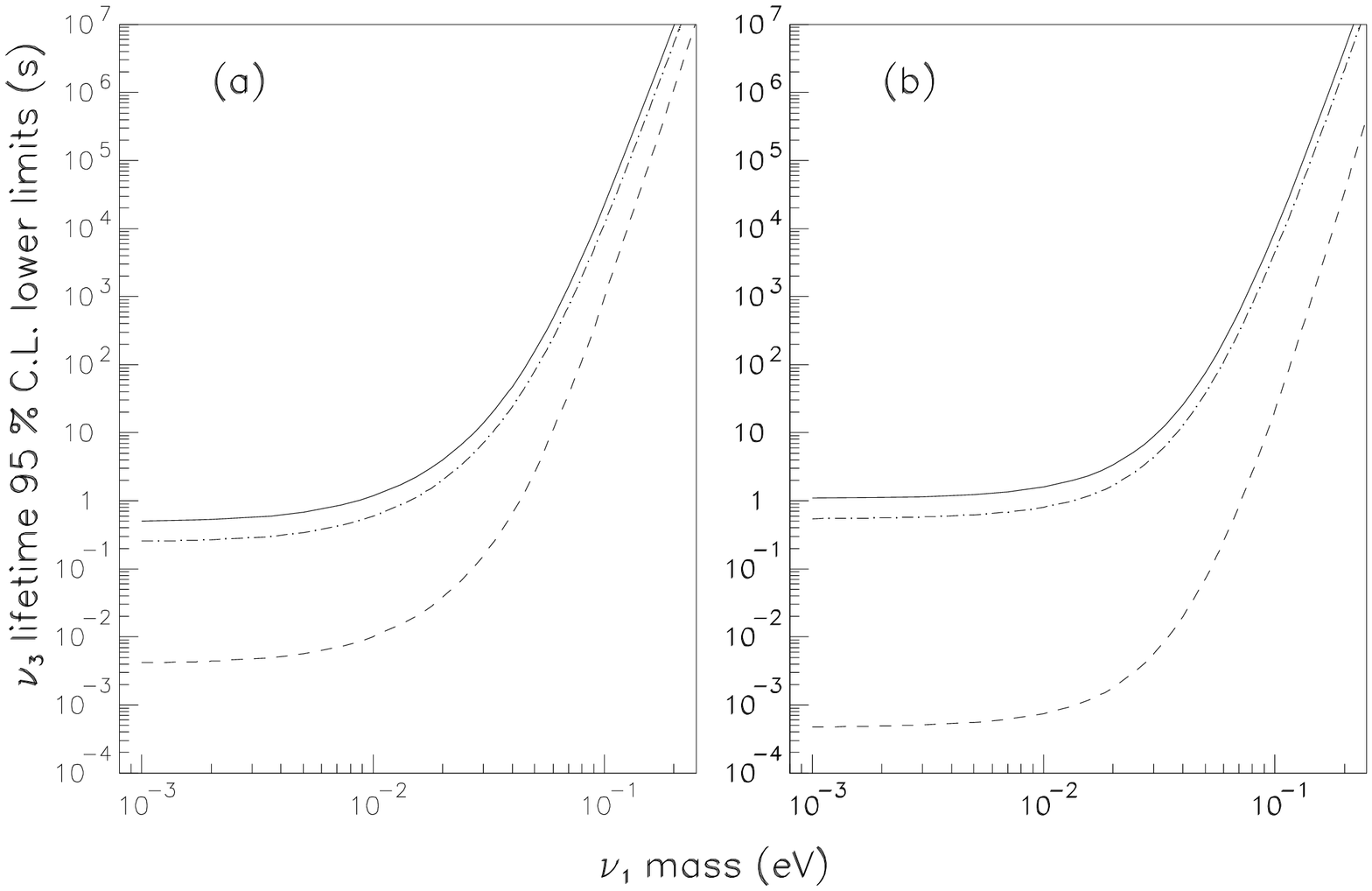}}
\caption{The 95\% CL lower limits for the $\nu_3$ proper lifetime,
as function of the $\nu_1$ mass, obtained from (a) set A (a) and (b) B.
 The solid, dot-dashed and dashed lines
correspond to three different neutrino polarizations, $\alpha = 1$, 0 and -1,
respectively.
 The results are obtained in a neutrino mixing scenario with three
generations, $\Delta m^2_{2,1} = 5 \times 10^{-5}$ eV$^2$ and $\Delta m^2_{3,(2,1)}
= 2.5 \times 10^{-3}$ eV$^2$.}
\end{center}
\end{figure}

\section{Conclusions}

We analyzed two sets of digital images obtained during the June 21$^{st}$ 2001
total solar
eclipse, in Zambia, looking for possible radiative decays of solar neutrinos, yielding
visible photons.

Data set A consists in a large number of frames  recorded
with a digital videocamera; it has a relatively large integration time, but
 a modest  space resolution.

Set B consists of 10  pictures taken with a
digital camera coupled to a small
telescope. Its time coverage is poorer than for set A,
but it has a better space resolution and
the instrument sensitivity was an order of magnitude better.

The proper lower lifetime limits
(95\% CL) obtained for the $\nu_2 \rightarrow \nu_1 + \gamma$
decays of lefthanded neutrinos range from
$\tau_0/m_2 \simeq 10$ s eV$^{-1}$ to  $\simeq
10^9$ s eV$^{-1}$, for $10^{-3}$ eV $< m_{\nu_1}
< 0.1$ eV, see Fig. 10.
 These limits are among the best
obtained from direct measurements, demonstrating the potentiality of
neutrino decay experiments during total solar eclipses (or possibly made in space, using
the Earth as light absorber \cite{frere}). The lab. lifetime limits
are about $10^7$ times
larger, thus the fraction of neutrino decays from the Sun to the Earth
would be negligeable.

A similar analysis was made for a possible $\nu_3 \rightarrow \nu_{2,1} + \gamma$ decay,
assuming $\sin^{2} \theta_{31} \simeq 0.1$ (the  value of
this mixing angle is not known).
 No signal compatible with a possible $\nu_3 \rightarrow \nu_{2,1} + \gamma$
is
seen.
The obtained 95\% C.L. $\nu_3$ proper lifetime lower limits, for $m_1 \geq 10^{-2}$ eV and
for $\alpha = -1$, 0,
are about two orders of magnitude
lower than for the $\nu_2$, Fig. 11.

 New observations,
 in better technical conditions,
during  forthcoming TSE's should be considered.

An attempt along these lines
 was made during the December 2002 eclipse, but the weather conditions
in South Africa did not allow any observation. We
intended to use three portable telescopes, equipped with astronomy type CCD's. The
sensitivity would have been
about two orders of magnitude better than what reported
in  this paper.

\section{Acknowledgments}

We would like to acknowledge many colleagues for useful comments and discussions.
We thank the
people of the Catania and Bologna Astronomical Observatories for their assistance during
calibrations.
Warm thanks are due to the Kiboko Safari, Lilongwe, Malawi, for their assistance
during the expedition in Zambia.
This work was funded by NATO Grant PST.CLG.977691 and partially supported by
the Italian Space Agency (ASI), INFN and the Romanian Space Agency (ROSA).

\end{document}